# Code Review Without Borders: Evaluating Synthetic vs. Real Data for Review Recommendation

Yogev Cohen
*School Of Computer Science, Faculty Of Sciences*
Holon Institute of Technology

Dudi Ohayon
*School Of Computer Science, Faculty Of Sciences*
Holon Institute of Technology

Romy Somkin
*School Of Computer Science, Faculty Of Sciences*
Holon Institute of Technology

Yehudit Aperstein
*Afeka Academic College of Engineering*
Tel Aviv Israel

Alexander Apartsin
*School Of Computer Science, Faculty Of Sciences*
Holon Institute of Technology

*Abstract*—Automating the decision of whether a code change requires manual review is vital for maintaining software quality in modern development workflows. However, the emergence of new programming languages and frameworks creates a critical bottleneck: while large volumes of unlabelled code are readily available, there is an insufficient amount of labelled data to train supervised models for review classification. We address this challenge by leveraging Large Language Models (LLMs) to translate code changes from well-resourced languages into equivalent changes in underrepresented or emerging languages, generating synthetic training data where labelled examples are scarce.

We assume that although LLMs have learned the syntax and semantics of new languages from available unlabelled code, they have yet to fully grasp which code changes are considered significant or review-worthy within these emerging ecosystems. To overcome this, we use LLMs to generate synthetic change examples and train supervised classifiers on them. We systematically compare the performance of these classifiers against models trained on real labelled data. Our experiments across multiple GitHub repositories and language pairs demonstrate that LLM-generated synthetic data can effectively bootstrap review recommendation systems, narrowing the performance gap even in low-resource settings. This approach provides a scalable pathway to extend automated code review capabilities to rapidly evolving technology stacks, even in the absence of annotated data.

## I. INTRODUCTION

As software ecosystems continue to diversify, the challenge of maintaining code quality across an increasingly complex range of languages and frameworks becomes more complicated. Automated tools that classify whether a code change requires manual review play a vital role in modern development workflows, improving both productivity and software reliability. However, such tools typically rely on supervised learning approaches that require substantial labelled datasets—resources that are often unavailable for newly emerging or less mature programming languages. While large repositories of unlabelled code exist for these languages, the lack of annotated examples limits the effectiveness of conventional machine learning methods.

Recent advancements in Large Language Models (LLMs) present a promising avenue for mitigating data scarcity. LLMs trained on vast corpora of code, such as GPT-4o, possess an impressive understanding of programming languages' syntax and semantics, even for newer or less popular languages. However, while these models can fluently generate code in these languages, they do not inherently understand which code changes are considered critical or review-worthy, knowledge typically learned from labelled examples.

In this work, we explore a novel transfer learning approach that leverages the generative capabilities of GPT-4 to synthesize labelled training data for low-resource languages. Specifically, we translate code changes from Java, where labelled review data is abundant, into equivalent C++ code changes using GPT-4. These synthetic examples are then used to train a review classification model based on CodeBERT, a transformer model pre-trained on source code (excluding C++ data).

We empirically evaluate this pipeline by comparing the performance of the classifier trained on LLM-generated synthetic C++ data with that of a baseline trained on real, labelled C++ data. Our experiments evaluate the effectiveness of synthetic data in approximating real-world review patterns and investigate the feasibility of utilizing LLM-driven data generation to extend automated code review systems to new languages and frameworks, thereby eliminating the need for costly manual annotation efforts.

## II. LITERATURE REVIEW

The emergence of large language models has spurred new approaches to automating code review tasks. For example, Lu et al. (2023) introduced **LLaMA-Reviewer**, an LLM-based framework fine-tuned with minimal data for code review assistance. Remarkably, even a 6.7 B-parameter LLaMA model, after parameter-efficient tuning, matched the performance of specialized code-review models. This demonstrates that pre-trained large language models (LLMs) can be adapted to generate helpful review comments without requiring massive retraining. In industry settings, LLM-driven review systems have already been deployed. Sun et al. (2025) describe BitsAI-CR. This two-stage code review pipeline utilizes fine-tuned large language models (LLMs) first to identify issues against a set of 219 review rules and then filter and verify the findings. Their system achieved high precision, with ~75% of suggestions being useful, and was rolled out to over 12,000 developers at ByteDance, demonstrating the practicality of LLM-based code analysis at scale. Notably, some approaches focus on classifying code changes rather than generating comments. For instance, Google's recent experimental method utilizes an LLM to classify issues in code modifications (e.g., flagging changes that violate a policy), rather than generating review text. This trend across research and industry underscores that LLMs can both generate review feedback and classify which code changes need attention, forming a foundation for automated review even when labelled examples are limited.

A key challenge in classifying code changes for new languages or frameworks is the lack of labelled training data. Recent work shows that transfer learning with LLMs can overcome this by leveraging knowledge from high-resource languages. Multilingual code models, such as CodeBERT (Feng et al., 2020), learn unified representations from source code in multiple languages, enabling some extent of zero-shot generalization. Empirically, Baltaji *et al.* (2024) conducted an extensive study of cross-lingual transfer on code tasks, finding that model performance can transfer surprisingly well between programming languages. In their experiments spanning 11–41 languages, models trained on one language (e.g., Kotlin or JavaScript, identified as particularly transferable sources) often performed well on others, even without target-language labels. Similarly, Li *et al.* (2022) demonstrated cross-lingual adaptation for a specific analysis task, namely type inference. They proposed **PLATO**, a framework that trains a model on one language's typed code and applies it to another language by focusing on standard syntactic features. PLATO significantly outperformed prior transfer-learning techniques – for example, using a model trained on Python to analyse JavaScript yielded over 5% absolute accuracy gain versus baseline methods. These results confirm that an LLM or transformer trained on one ecosystem can be repurposed for another, which is crucial for classifying or reviewing code in new languages. Through transfer learning, an LLM could learn what kinds of changes are risky or require review from one language's data and then apply that knowledge to a novel language where such data is unavailable.

Another approach to handling unlabelled code changes in a new language is automatic code translation using large language models (LLMs). By translating an unfamiliar-language code change into a language where tools or classifiers exist, one can leverage existing review knowledge. Research in the last few years has made great strides in LLM-based code translation. Rozière *et al.* (2020) achieved a breakthrough with **TransCoder**, an unsupervised transformer model that learned to translate code among C++, Java, and Python without any parallel training data. Their approach combined cross-lingual language modelling and back-translation, yielding strong results on this challenging task. This demonstrated that an LLM can automatically learn semantic mappings between languages – a form of cross-language generalization – purely from large monolingual code corpora.

Subsequent models incorporated code translation into their training objectives. Ahmad *et al.* (2021) introduced **PLBART**, a sequence-to-sequence transformer pre-trained on Java and Python functions, along with their documentation, via denoising autoencoding. PLBART can perform code generation and translation in *multiple* languages; indeed, it outperformed or matched state-of-the-art methods on code-to-code translation benchmarks covering seven programming languages. In a similar vein, Wang *et al.* (2021) developed **CodeT5**, which leverages T5-style text-to-text training on diverse code. These multilingual code language models (LLMs) effectively learn a shared space for different programming languages. As a result, they enable translating a code change written in an "unknown" language to a known one, or even directly assessing the change in a language-agnostic way. Such capabilities are highly relevant for cross-language code review: an LLM could, for example, take a code diff in a new framework and explain or classify it by internally mapping it to concepts learned from other languages.

When real labelled data is scarce, generating synthetic data has emerged as a powerful strategy for training and adapting models. Recent studies have leveraged large language models (LLMs) to create artificial code examples or annotations that serve as substitutes for human-labelled data. For instance, Zhu *et al.* (2024) present an approach called **MIRACLE** to improve code change translation through synthetic data. MIRACLE utilizes a pre-trained code model to translate functions from one language to another, and employs static analysis and compilation checks to curate high-quality parallel code pairs. By adding thousands of these LLM-generated pairs to the training set, they significantly enhanced translation accuracy for low-resource language pairs, even outperforming code-specific LLMs that were 10 times larger in parameters. Notably, their method achieved up to a 43% improvement in C code translation despite having fewer than 150 real examples for that case. This illustrates how synthetic diffs or code pairs can fill in gaps in training data. In the realm of code review and defect detection, synthetic data can be created by *simulating* code changes. Allamanis *et al.* (2021) demonstrated this with **BugLab**, a self-supervised system for bug finding. BugLab trains a *bug selector* model to insert bugs into code deliberately and a *bug detector* model to catch them, effectively letting the model "learn from its own mistakes". This hide-and-seek training yielded a 30% improvement over baseline methods on real-world bug datasets, all without any manually labeled examples of buggy code. The success of BugLab shows that an ML model can learn to flag problematic code changes (in this case, buggy edits) by training on synthetically generated changes.

More broadly, researchers are recognizing that LLMs can be utilized to synthesize data for various software engineering tasks. As noted by Nadas and Diosan (2025) in their recent survey, LLM-generated code data (from synthetic code review examples to machine-generated coding tasks) has proven effective for augmenting low-resource scenarios. The ability to produce diverse code snippets, translations, or commit messages on demand means an LLM can supply a pseudo-labelled dataset for a new language or framework where one did not previously exist. This synthetic data, when carefully filtered for accuracy, can then be used to supervise a model in determining which code changes require review. In summary, the literature indicates a convergence of transfer learning, cross-language large language models (LLMs), and synthetic data generation as enabling techniques. Together, these advances allow LLM-based classifiers to determine whether a given code change requires human review, even in novel programming languages or ecosystems, by leveraging generalized knowledge learned from other contexts and supplemented with artificial training examples.

### III. METHODOLOGY

We generate a labeled C++ code review dataset by translating labeled Java code changes from the dataset introduced in *Automating Code Review Activities by Large-Scale Pre-training*. Each Java code change includes a binary label indicating whether it required manual review. Using GPT-4o, we translate these Java changes into C++ while preserving the original intent and the assigned review label. GPT-4o is prompted to produce functionally equivalent C++ code changes that reflect the same significance as the original Java changes. We assume that while GPT-4o understands the syntax and semantics of C++, it does not inherently know which changes are critical, so the original review label is retained for the translated code. Basic validation through static analysis ensures that the generated C++ code is syntactically correct. The resulting synthetic dataset enables the training of a CodeBERT classifier to predict review needs for C++ code changes, addressing the lack of labeled data for this language.

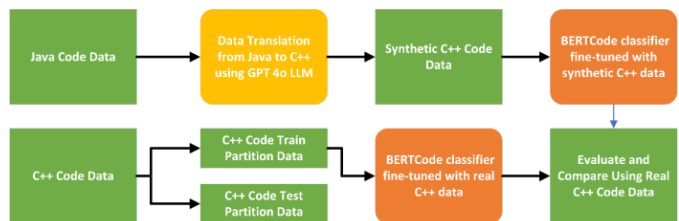

*Figure 1: Data generation, training, and evaluation flow*

Additionally, we extract the original C++ code changes from the source dataset and partition them into training and testing sets. The training portion is used to create a baseline classifier using real C++ examples. In contrast, the test portion is used to evaluate both the baseline model and the model trained on synthetic C++ examples generated by GPT-4o. This setup allows a direct comparison of model performance using real versus synthetic training data.

We use a CodeBERT-based classifier to predict whether a C++ code change requires manual review. For the baseline case, the model is fine-tuned on the real C++ training data extracted from the source dataset. For the synthetic case, the same model architecture is fine-tuned using the C++ code changes generated by GPT-4o from the labeled Java examples. In both cases, the test set, comprising fundamental C++ code changes not encountered during training, is used for evaluation. This ensures that the performance comparison between models trained on real and synthetic data is conducted under identical conditions. Evaluation metrics such as accuracy, precision, recall, and F1-score are computed on the same test set to assess the effectiveness of using synthetic data for code review classification.

### IV. RESULTS

We evaluate the performance of the CodeBERT-based classifier in two training scenarios: using real C++ code review data and using synthetic C++ examples generated from Java code changes via GPT-4o. Both models are evaluated on the same held-out real C++ test set to ensure fair comparison.

| Training Data | Accuracy | Precision | Recall | F1 |
|---|---|---|---|---|
| Real C++ | 0.65 | 0.64 | 0.65 | 0.64 |
| Synthetic C++ | 0.65 | 0.65 | 0.68 | 0.66 |

**Table 1**: Evaluation results on real and synthetic data

The results show that the model trained on real C++ data achieves higher performance across all evaluation metrics. However, the model trained on synthetic data demonstrates competitive results, indicating that LLM-generated training data can effectively approximate real-world review patterns in low-resource scenarios

### V. CONCLUSIONS AND FUTURE RESEARCH

This work demonstrates the feasibility of using large language models, specifically GPT-4o, to generate synthetic code review datasets for low-resource programming languages. By translating labeled Java code changes into C++ while preserving review labels, we enable supervised training of a CodeBERT-based classifier in the absence of sufficient real-world, C++-labeled data. Although models trained on real data still outperform those trained on synthetic data, the performance gap is narrow enough to suggest that LLM-generated examples can be a practical solution when real data is unavailable.

Future work will explore extending this methodology to additional programming languages and frameworks, particularly those emerging in specialized domains such as mobile development (e.g., Swift, Kotlin) and data science (e.g., Julia, R). We also plan to investigate the use of more advanced prompting techniques and reinforcement learning strategies to improve the quality and realism of the synthetic code changes. Additionally, incorporating domain-specific review criteria and experimenting with multilingual code models may further enhance cross-language transferability and review recommendation accuracy.

.